\documentstyle [12pt]{article}
\begin{document}
\title{A generalization of Jaynes' principle: \\
       an information-theoretic interpretation\\
       of the minimum principles of\\
       quantum mechanics and gravitation}

\author{John H. Van Drie\\
        www.johnvandrie.com \\
        Kalamazoo, MI 49008  USA}

\maketitle

\begin{abstract}
By considering the "kinetic-energy" term of the minimum principle for the
Schr\"{o}dinger equation as a measure of information, that minimum principle
is viewed as a statistical estimation procedure, analogous to the manner in which
Jaynes ({\it Phys. Rev.},{\bf 106}, 620, 1957) interpreted statistical mechanics.
It is shown that the entropy formula of Boltzmann and Jaynes obey a property in
common with the quantum-mechanical kinetic energy, in which both quantities are
interpreted as measures of correlation.  It is shown that this property is shared
by the key terms in the minimum principles of relativistic quantum mechanics
and General Relativity.  It is shown how this principle may be extended to non-Riemannian
nonEuclidean spaces, which leads to novel field equations for the torsion.
\end{abstract}

\section{INTRODUCTION}

As Laplace famously observed, Newtonian mechanics is deterministic to an
idealized intelligent being, but (as Laplace observed less-famously in the same
passage) practical realities demand a probabilistic mechanics \footnote{Laplace,
P.-S., Laplace's Th\'{e}orie Analytique des Probabilit\'{e}s, 3rd ed., Paris: Courcier, 
1820.  Reprinted in volume 7 of his Oeuvres Compl\`{e}tes, Paris: Gauthier-Villars, 
1886, see the Introduction, pp. vi-ix.}.  In the century after 
Laplace's observation, two forms of probabilistic
mechanics were discovered:   statistical mechanics, and quantum mechanics.
While the former maintained at a detailed level the determinism of Newtonian
mechanics, acquiring its probabilistic nature only when describing macroscopic
observations of large ensembles of particles, quantum mechanics introduced a
probabilistic nature at the most fundamental level.

In a sharpening of Laplace's "principle of insufficient reason", Jaynes cast
statistical mechanics into a novel form.  Jaynes equated the entropy of
statistical mechanics

\begin{equation}
- \sum_i p_i \log(p_i)
\end{equation}
to the entropy of Shannon's information theory, and equated the principle of
maximum entropy with an information-theoretic inference law, Jaynes' principle,
which asserts that the maximum entropy probability distribution is that
distribution which is the least-biased estimate for such a distribution 
\footnote{Jaynes, E.T., {\it Phys. Rev.},{\bf 106},620 (1957)}.

The goal here is to generalize Jaynes' principle to the continuous probability
distributions of quantum mechanics, and to demonstrate how the extremum
principles of quantum mechanics, the Schr\"{o}dinger equation and the Dirac
equation, may be viewed as statements for formulating least-biased estimates of
those continuous probability distributions.  My thesis \footnote{Van Drie, J.H.,
thesis,Calif. Inst. of Tech., 1978, unpublished}
explored the conceptual implications of this view.  In this work, we focus on
the surprising consequence of this perspective that the laws of gravitation as
defined by General Relativity may also be viewed as a minimum-information
principle, and show how this leads naturally to a set of field equations for
non-Riemannian nonEuclidean geometries.

\section{ENTROPY AS A MEASURE OF CORRELATION}

As pointed out in an earlier work \footnote{Van Drie, J.H., http://xxx.lanl.gov/abs/math-ph/0001024}, 
the Boltzmann / Shannon
formula for the entropy of a discrete distribution may be viewed as a measure of
the correlation between two distinct distributions.  This was demonstrated, by
showing that the entropy functional is non-decreasing as "correlation-destroying
transformations" are applied to the distribution.  This perspective yields yet
another view of Jaynes' principle:  the least-biased distribution is that which
displays the least correlation between the variables consistent with the
constraints.  Also, this perspective leads to yet another view of the Second Law
of thermodynamics:  correlations tend to spontaneously decay, and are highly
unlikely to arise spontaneously in an isolated system.

It is useful to recall the fundamental importance that the concept of
correlation played in Maxwell's original derivation of the velocity distribution
of the atoms of an ideal gas \footnote{Maxwell,J.C., {\it Phil. Soc.}, 1860}.  
He placed two assumptions on
the velocity distribution $\Phi ( v_x, v_y, v_z)d^3v$:  (1) there should be no
preferred orientation, $\Phi (v_x, v_y, v_z)d^3v = \Phi (v) d^3v$, where $v^2 =
v_x^2+v_y^2+v_z^2$, and (2) the velocities along one direction should not be
correlated with those along another direction, 
\begin{equation}
\Phi(v_x, v_y, v_z)dv_x dv_y dv_z = \phi(v_x)dv_x \phi(v_y)dv_y \phi(v_z)dv_z
\end{equation}
These assumptions lead to Maxwell's velocity distribution law, 
\begin{equation}
\Phi(v_x, v_y, v_z)d^3v = exp(-\alpha v^2) v^2 dv d\Omega
\end{equation}
where $\alpha$ is a positive constant (shown by Boltzmann to be
1/kT), and $d\Omega$ is the element of surface integration in velocity space,
$\sin(v_\theta) dv_{\theta} dv_{\phi}$.  The ability of a distribution over multiple
variables to be expressed as the product of distributions over single variables
is the hallmark of an uncorrelated distribution.

\section{KINETIC ENERGY AS A CORRELATION MEASURE}

The term in the Hamiltonian associated with the Schr\"{o}dinger equation, $<\Psi |
\Delta | \Psi>$ is commonly called the "kinetic energy" (multiplied by a suitable
units-dependent constant), by analogy to the corresponding term in the classical
Hamiltonian, where $\Delta$ is the Laplacian operator, 
\begin{equation}
\Delta =  -
\frac{\partial^2}{\partial x^2} - \frac{\partial^2}{\partial y^2} -
\frac{\partial^2}{\partial z^2}
\end{equation}

It was originally suggested by this author in unpublished work 
\footnote{Van Drie, J.H., Candidacy report, Calif. Inst. of Tech., 1975}, and 
later by Sears, Dinur and Parr 
\footnote{Sears, S., Parr, R., Dinur, U., {\it Isr. J. Chem.}, {\bf 19}, 165 (1980)} 
that this expression represents an entropy expression.  
This assertion rested on intuitive arguments, leaving open the question "what
mathematical property is common to both the kinetic energy of quantum mechanics
and the entropy of statistical mechanics?".  The answer which will be provided
here is that both expressions are quantitative measures of correlation, or the
lack thereof; the idea that the quantum mechanical kinetic energy in some
instances measures correlation is an old concept from molecular quantum
mechanics \footnote{L\"{o}wdin,P.O., {\it Adv. Chem. Phys.}, {\bf 2},207, (1959)}.

 Let us consider two spaces $M_1$ and $M_2$, of dimension $n_1$ and $n_2$ respectively,
and the cross-product space $M_1 \times M_2$ of dimension $n_1 + n_2$.   Furthermore, let
us consider a representation of the group of rotations and translations on $M_1,
\gamma_1$, and a representation on $M_2, \gamma_2$.  As in quantum mechanics, let
us assume that a metric exists which allows us to define a probability
distribution over $M_k$ from $\gamma_k$,  $\rho_k = ( \gamma_k, \gamma_k )$ for each
point in $M_k$.  One can construct the product representation, $\gamma_1 \times
\gamma_2$, as a representation over $M_1 \times M_2$.   Such a representation on $M_1 \times 
M_2$ is by definition uncorrelated relative to the variables $M_1$ vis-\`{a}-vis $M_2$,
since it can be written as the product of a representation on $M_1$ and one on
$M_2$.

Consider an operator on representations of $M_1, O_1$, and an operator on
representations of $M_2$, and the composition of these on $M_1 \times M_2$, $O_{1 \times 2}$.  We
assert that all such operators which obey the following relationship may be
considered as measures of correlation between the variables of $M_1$ and those of
$M_2$:

\begin{equation}
    O_{1 \times 2} = O_1 + O_2  \label{eq:Master}
\end{equation}

Denoting the expectation value of an operator $O_k$ over the space $M_k$  against
the representation $\gamma_k$ by \linebreak
$<\gamma_k, O_k \gamma_k>$, which equals

\begin{equation}
  <\gamma_k, O_k \gamma_k > = \int_{M_k} (\gamma_k, O_k \gamma_k)
\end{equation}
we see that for operators which are considered measures of correlation against
$M_1$ and $M_2$, i.e. those operators obeying the relation (~\ref{eq:Master}),

\begin{equation}
<\gamma_1 \times \gamma_2, O_{1 \times 2} \gamma_1 \times \gamma_2 > = <\gamma_1 O_1 \gamma_1> +
<\gamma_2 O_2 \gamma_2>
\end{equation}

Considering these expectation values as measures of correlation of their
corresponding representations, denoted $I[\gamma]$, this allows us to interpret
the above equation to say that, for uncorrelated representations, the measure of
correlation is additive:

\begin{equation}
I[\gamma_1 \times \gamma_2] = I[\gamma_1] + I[\gamma_2]
\end{equation}

This same relationship holds for the Boltzmann / Shannon entropy of two discrete
distributions, $P = \{ p_i \}_{i=1}^n$ and $Q = \{ q_j \}_{j=1}^m$, where the product
distribution $PQ = \{ p_i q_j \}_{i,j=1}^{n,m}$,

\begin{equation}
S(PQ) = S(P) + S(Q)
\end{equation}

Note that the Laplacian operator, the kinetic energy term of the minimum
principle for the Schr\"{o}dinger equation, explicitly obeys the property (~\ref{eq:Master})
for the n translational variables $\{ x_k \}_{k=1}^{n}$ of n-dimensonal space M, since
for any m-dimensional subspace $\{ x_k \}_{k=1}^m M_1$, and  its complement
$\{ x_k \}_{k=m+1}^n M_2$, $\Delta_M = \Delta_{M_1} + \Delta_{M_2}$.

Intriguingly, R. A. Fisher applied the term information to the
expectation value of the Laplacian, in an obscure and unelaborated reference
\footnote{Fisher, R.A., {\it Stat. Methods and Sci. Inference}, 
NY:  Hafner, 1956, eqn. 155,
but in the Introduction, he maintains this has no relation to the concepts
in the Mathematical Theory of Communication, i.e. information theory}.

Even for 4-component spinor representations in relativistic spacetime, where the
relevant quantum mechanical equation is the Dirac equation and the corresponding
term in the minimum principle is the expectation value of the operator
$i/\hspace{-0.1in}\nabla$ \footnote{Feynman R.P., {\it Quantum Electrodynamics}, London: 
Benjamin/Cummings, 1962}, the property (~\ref{eq:Master}) holds.   For the flat
spacetime metric $g_{\mu \nu}$, a set of 4x4 matrices defined over the
components of the spinors exist which obey the property

\begin{equation}
   \gamma_{(\mu} \gamma_{\nu)} = \frac{1}{2} ( \gamma_\mu \gamma_\nu +
\gamma_\nu \gamma_\mu) = g_{\mu \nu}
\end{equation}
and the operator $i/\hspace{-0.1in}\nabla$ may be written

\begin{equation}
   i/\hspace{-0.1in}\nabla= i \sum_\mu \gamma_\mu \frac{\partial}{\partial x^{\mu}}
\end{equation}

As with the Laplacian, the property (~\ref{eq:Master}) is evident from the definition of
this operator; hence even in relativistic spacetime, the minimum principle
contains a term which we may call a measure of correlation.

\section{THE GENERALIZED JAYNES' PRINCIPLE}

Jaynes asserted that maximizing the entropy $-\sum_k p_k \log(p_k)$ over all
distributions ${p_k}$ subject to the constraints of a given energy $E = \sum E_k
p_k$ and normalization $\sum_k p_k = 1$ may be viewed as a statement that the
distribution $p_k$ is the least-biased distribution for ${p_k}$ subject to these
constraints.

\begin{eqnarray}
   \delta_{p_k} \{ -\sum_k p_k log(p_k) & - & \lambda_1 (\sum_k E_k p_k) \\
                                & - & \lambda_2 (\sum_k p_k) \} = 0\\
\Longrightarrow p_k & = & \alpha exp(-\beta E_k)
\end{eqnarray}
where $\lambda_1$ and $\lambda_2$ are Lagrangian multipliers, and $\delta_{p_k}$ denotes
varying over all possible $\{p_k\}$.  Viewing the Laplacian
operator as a measure of correlation in a representation $\gamma$, we assert that
the minimum principle for the Schr\"{o}dinger equation may be viewed as the
statement that $\gamma$ represents the least-biased representation subject to the
constraints of normalization and in the presence of a potential V(x):

\begin{equation}
   \delta_{\gamma} \{ < \gamma, \Delta \gamma > - \lambda_1 \int V(x) (\gamma, \gamma) 
-\lambda_2 \int (\gamma, \gamma) \} = 0
\end{equation}

One of the many conceptual implications of this view is that it allows us to
understand the physical basis for why the electron does not collapse onto the
nucleus of an atom:  the tendency to minimize its potential energy by
withdrawing into the nucleus is counterbalanced by the tendency of its
distribution to resist achieving such a highly-correlated state.

An amusing application of this principle is the case where $\gamma$ is a vector
representation, the normal to the surface of a soap-film.  Minimizing $<\gamma,
\Delta \gamma>$ subject to the constraint that the film adhere to a specified
1-dimensional wire frame gives the equation for the equilibrium configuration of
such surfaces.   This tendency of minimizing $<\gamma, \Delta \gamma>$ to
function like a surface tension can also be understood by recalling Maxwell's
observation about the Laplacian:  the Laplacian of a function is proportional to
the difference between that function and that function's average value over a
ball of radius $\epsilon$ \footnote{Maxwell, J.C., {\it Treatise on Electricity and 
Magnetism, vol. 1}, New York:  Dover, 1991, p. 31}, a property well-known in the
numerical analysis of Laplace's equation.

For nonEuclidean geometries, a more refined definition of the Laplacian must be
used to ensure that the property (~\ref{eq:Master}) is satisfied.  The generalized
Laplacian of deRham, 
\footnote{deRham, G.,
Vari\'{e}t\'{e}s Diff\'{e}rentiables, Paris: Hermann, 1960, p. 125}, 
$\Delta = d \delta  + \delta d$, 
where d is the exterior derivative and $\delta = \ast d \ast$, $\ast$ the Hodge dual
operator, may be tediously shown to possess the property (~\ref{eq:Master}), through the
use of deRham's {\it forms double}.  This allows us to write the most general form of
this generalized Jaynes' principle, namely

\begin{equation}
   \delta_{\gamma} \{ <\gamma, \Delta \gamma> \} = 0 
\end{equation}
subject to constraints, among them $<\gamma,\gamma> = 1$,
where $\gamma$ is understood to be any representation over a nonEuclidean space,
and $\Delta$ is understood to be deRham's Laplacian.

\section{APPLICATION TO NONEUCLIDEAN GEOMETRIES}

For a Riemannian geometry, deRham gave the explicit formula for his Laplacian
applied to a tensor of arbitrary rank p\footnote{deRham,{\it ibid.}, p. 131}:

\begin{eqnarray} 
\Delta \alpha_{k_1 k_2 \dots k_p} & = & -  \alpha_{k_1 k_2 \dots k_p;i}^{;i} +
\sum_{\nu=1}^p (-1)^{\nu} R^{h.i.}_{.i.k_{\nu}} \alpha_{h k_1 \dots
\widehat{k_{\nu}} \dots k_p} \\
&& + 2 \sum_{\mu < \nu}^{1 \dots p} (-1)^{\mu + \nu} R^{h.i.}_{.k_{\nu}.k_{\mu}}
\alpha_{ihk_1 \dots \widehat{k_{\mu}} \dots \widehat{k_{\nu}} \dots k_p}
\end{eqnarray}
where $\widehat{k_{\mu}}$ denotes that subscript is dropped from the enumerated 
list of indices, and where deRham's notation of the covariant derivative 
$\nabla_i$ is replaced by the notation of Misner et al.\footnote{Misner, C.W., 
Thorne, K.S., and Wheeler, J.A., {\it Gravitation}, San Francisco:  W.H.Freeman, 
1973}where the covariant derivative is denoted by ${;\alpha}$, and where 
$R^{\alpha}_{\beta \mu \nu}$ is the Riemann curvature tensor.  
In spacetime, applying deRham's Laplacian to the metric,  $g_{\mu \nu}$, 
we see that the above formula reduces to

\begin{equation}
  \Delta g_{\mu \nu} = g_{\mu \nu; \alpha}^{\hspace{0.18in};\alpha} + R_{\mu \nu}
\end{equation}
where $R_{mu \nu}$ is the Ricci tensor, the contraction of the Riemann tensor.
The term $g_{\mu \nu; \alpha}^{\hspace{0.1in};\alpha}$ is zero, by the covariant constancy of
the metric, and hence the measure of correlation, the expectation value of the
Laplacian, is

\begin{equation}
  < g^{\mu \nu} \Delta g_{\mu \nu} > = \int g^{\mu \nu} R_{\mu \nu} d\tau = \int R
d\tau
\end{equation}
where $d\tau$ is the volume element of integration over spacetime, and R is the
scalar curvature, the contraction of the Ricci tensor.  Minimizing $\int R d\tau$
over all metrics is the Hilbert variational principle, and leads to Einstein's
equations of General Relativity in free space\footnote{Misner {\it et al.}, 
\it{ibid.}, Ch. 21}.

Hence, the generalized Jaynes principle states that Einstein's equations of
General Relativity in free space may be interpreted as asserting that the metric
is the least-biased metric, or the minimally-correlated metric.

In the nonEuclidean spaces of Cartan, the fundamental quantities are not the 
metric, but rather the 1-forms of the repere mobile $\omega_{\mu}$ and the 
connection 1-forms $\omega^{\mu}_{\nu}$\hspace{0.1in}\footnote{Slebodzinski, W., {\it Exterior 
Forms and their Applications}, Warsaw:  Polish Scientific Publishers, 1970, 
section 124.  This originated from Cartan, E., {\it Ann. Ec. Norm.},{\bf 40}, 325 
(1923), reprinted in his Oeuvres Compl\`{e}tes, partie III, vol. 1, Paris: 
Gauthiers-Villars, p. 659}.  He showed that, for these more general
nonEuclidean spaces, an additional invariant arises, the torsion; two such
spaces are equivalent if both the torsion and curvature are equal.  Cartan's
structure equations define the torsion $\Omega^{\mu}$ and curvature
$\Omega^{\mu}_{\nu}$:

\begin{eqnarray}
    d \omega^{\mu} + \omega^{\mu}_{\nu} \wedge \omega^{\nu} & = & \Omega^{\mu} \\
    d \omega^{\mu}_{\nu} + \omega^{\mu}_{\alpha} \wedge \omega^{\alpha}_{\nu} & = &
\Omega^{\mu}_{\nu}
\end{eqnarray}

For Riemannian spaces, the torsion is zero, and the problem of equivalence
reduces to the study of the curvature form.  While Einstein's equations of
General Relativity allows one to write field equations for the curvature, the
issue of field equations for torsion and curvature has received less attention.

The natural extension of the above ideas to such spaces with torsion is to
consider the following equations:

\begin{equation}
 \delta_{\omega_{\mu},\omega^{\mu}_{\nu}} \{    < \omega^{\mu} \Delta \omega_{\mu} > \}  = 0
\end{equation}
where the minimization is taken over all orthonormal bases $\omega_{\mu}$ and
connection forms $\omega^{\mu}_{\nu}$.
Of the numerous possible forms to choose from, the basis 1-forms seems the most
natural, in that, like the metric, they represent a generalized potential,
derivatives of which lead to a generalized force; derivatives of generalized
forces can then be linked to source terms, like mass/energy density.  The
details of the above equation will be explored in a future work.

\section{IMPLICATIONS OF THIS EQUATION}
  
An interesting question is ``Are the principles of structure independent of 
scale?''.  One of the reasons the study of fractals in biological settings 
has generated such enthusiasm is that it implicitly answers that question 
Yes over the scales ranging from the size of macromolecules to the size of 
plants.  This equation appears to suggest that the principles of structure 
may be independent of scale over an even wider range, from the atomic scale 
to the astrophysical scale.  Of course, at each scale the forces that are 
relevant are different, and hence the resulting structures are different 
(the constraints that must be imposed in the generalized Jaynes principle).
\end{document}